\documentclass{article}
\usepackage{spconf,amsmath,amssymb,graphicx}
\usepackage{graphicx}
\usepackage{subcaption}

\title{A Novel Approach to WaveNet Architecture for RF Signal Separation with Learnable Dilation and Data Augmentation}

\name{Yu Tian$^{1}$, Ahmed Alhammadi$^{1}$, Abdullah Quran$^{2}$, Abubakar Sani Ali$^{2}$}
\address{$^{1}$Technology Innovation Institute, 9639 Masdar City, Abu Dhabi, UAE \\
         $^{2}$Khalifa University, Abu Dhabi, UAE}

\begin{document}
%
\maketitle
\begin{abstract}
In this paper, we address the intricate issue of RF signal separation by presenting a novel adaptation of the WaveNet architecture that introduces learnable dilation parameters, significantly enhancing signal separation in dense RF spectrums. Our focused architectural refinements and innovative data augmentation strategies have markedly improved the model's ability to discern complex signal sources. This paper details our comprehensive methodology, including the refined model architecture, data preparation techniques, and the strategic training strategy that have been pivotal to our success. The efficacy of our approach is evidenced by the substantial improvements recorded: a 58.82\% increase in SINR at a BER of $10^{-3}$ for OFDM-QPSK with EMI Signal 1, surpassing traditional benchmarks. Notably, our model achieved first place in the challenge \cite{datadrivenrf2024}, demonstrating its superior performance and establishing a new standard for machine learning applications within the RF communications domain.
\end{abstract}
\begin{keywords}
Radio Frequency Signal Separation, Machine Learning, WaveNet Architecture, Learnable Dilation, Data Augmentation
\end{keywords}

\section{Introduction}
\label{sec:intro}

The co-channel signal separation in the crowded radio-frequency (RF) spectrum is a crucial task for enabling various wireless systems to operate simultaneously. Our research, which we submitted for the MIT RF Challenge \cite{datadrivenrf2024}, introduces a machine learning algorithm specifically designed to isolate a target signal from a blend of overlapping interfering signals. We explore the question: How can deep learning be harnessed to overcome this signal separation challenge?

\section{Methodology}
\label{sec:methodology}
In this section, we discuss our modified WaveNet architecture with learnable dilation and the optimization of our dataset for the RF challenge \cite{datadrivenrf2024}. We detail our model-related enhancements and then outline our data-related contributions, both crucial for improving RF signal separation performance.

\subsection{Model-Related Contributions}

\begin{figure}[t]
\centering
\includegraphics[width=1\linewidth]{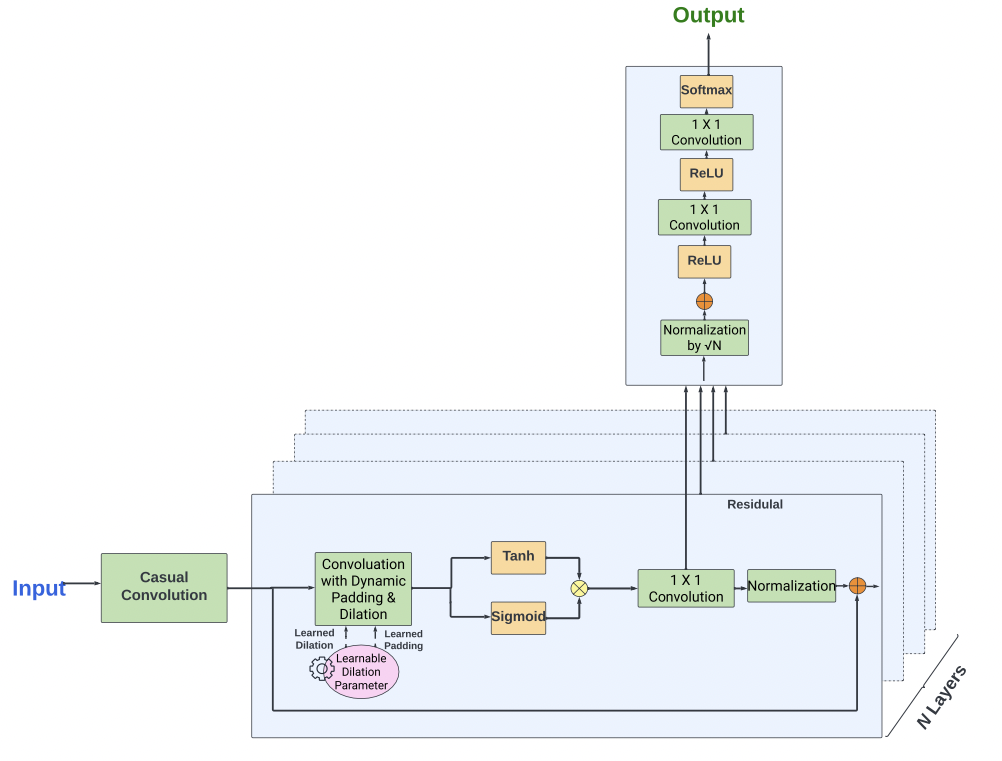}
\caption{Modified Wavenet with Learnable Dilation and Padding} \label{fig_1}
\vspace{-2mm}
\end{figure}

\begin{figure}[t]
\centering
\includegraphics[width=0.8\linewidth]{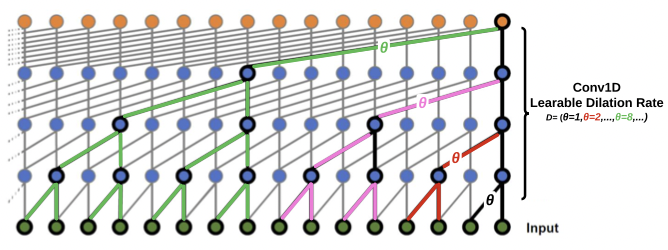}
\caption{An Illustration of Learnable Dilation Rate} \label{fig_2}
\vspace{-2mm}
\end{figure}

In adapting WaveNet for RF signal separation, significant modifications based on the benchmark, as shown in Fig. \ref{fig_1}, were made to suit the specific needs of RF signal processing. The introduction of learnable dilation parameters within the convolutional layers stands as a key enhancement. This allows the model to adaptively modulate its receptive field, focusing on the most relevant features of RF signals for effective separation, as shown in Fig. \ref{fig_2}.
Additionally, hyperparameter optimization was a critical part of our enhancement process. We increased the number of residual channels from 128 to 256, allowing for a richer representation of the signals. The dilation cycle length was also customized to the specific type of RF mixture being processed. These adjustments were crucial in enhancing the model's ability to capture and process the temporal structure of the signals, significantly boosting its performance in RF signal separation tasks.

\subsection{Data-Related Contributions}
Recognizing the pivotal role of training data, we focused on improving the quality and nature of the data used in our experiments. Our data-related contributions are as follows:

\begin{itemize}
    \item \textbf{Validation Data Set Reconfiguration:} We diverged from the conventional approach of using a partition of the training data as the validation set. Instead, we employed the test data provided by the challenge organizers for validation purposes. This ensured that our model was evaluated against a more robust and diverse set of examples, for enhanced generalizability.
    \item \textbf{Data Augmentation for CommSignal2:} In an innovative twist, we generated additional training points for CommSignal2 by converting high SNR waveforms with zero BER into bits and then reconverting these bits back into waveforms. This technique effectively increased the diversity and size of our training dataset.
\end{itemize}
\section{Data Preparation and Model Training}

The dataset includes 1100 segments, each with 100 examples of 40960-sample long signal estimates at 11 different SINR levels, along with corresponding bit string estimates critical for the demodulation process. It is split into training and validation sets—the former to tune model parameters and the latter to evaluate generalizability. 
With regards to the model training, we adopt a supervised learning approach, aiming to minimize the mean squared error (MSE) between the model's predictions and the actual signal sources. The model iterates through epochs, with periodic validation checks to fine-tune the learning rate and determine if early stopping is necessary. Additionally, we generated 22000 examples of augmented data for CommSignal2.

\section{Results}
Fig. 3 illustrates the superior performance of our method across four scenarios. Subfigures (a) and (b) depict the Bit Error Rate (BER) for OFDM-QPSK with EMI Signal 1 and QPSK with Comm Signal 2, respectively. Our method demonstrates a consistently lower BER throughout the entire SINR range. For OFDM-QPSK with EMI Signal 1, our method achieves a BER of \(10^{-3}\) at an SINR that is \(10 \, \text{dB}\) lower than the Default\_Torch\_WaveNet, reflecting a \(58.82\%\) improvement in SINR performance at this BER level. For QPSK with Comm Signal 2, our method reaches the same BER at an SINR that is approximately \(5 \, \text{dB}\) lower, which translates to an improvement of approximately \(33.33\%\). Subfigures (c) and (d) present the Mean Squared Error (MSE) for the same signal configurations, further confirming our method's enhanced accuracy in signal reconstruction, with a notable decrease in MSE values across the SINR spectrum.

\label{sec:pagestyle}
\begin{figure}[t]
    \centering
    \begin{subfigure}[b]{0.45\linewidth}
        \includegraphics[width=\linewidth]{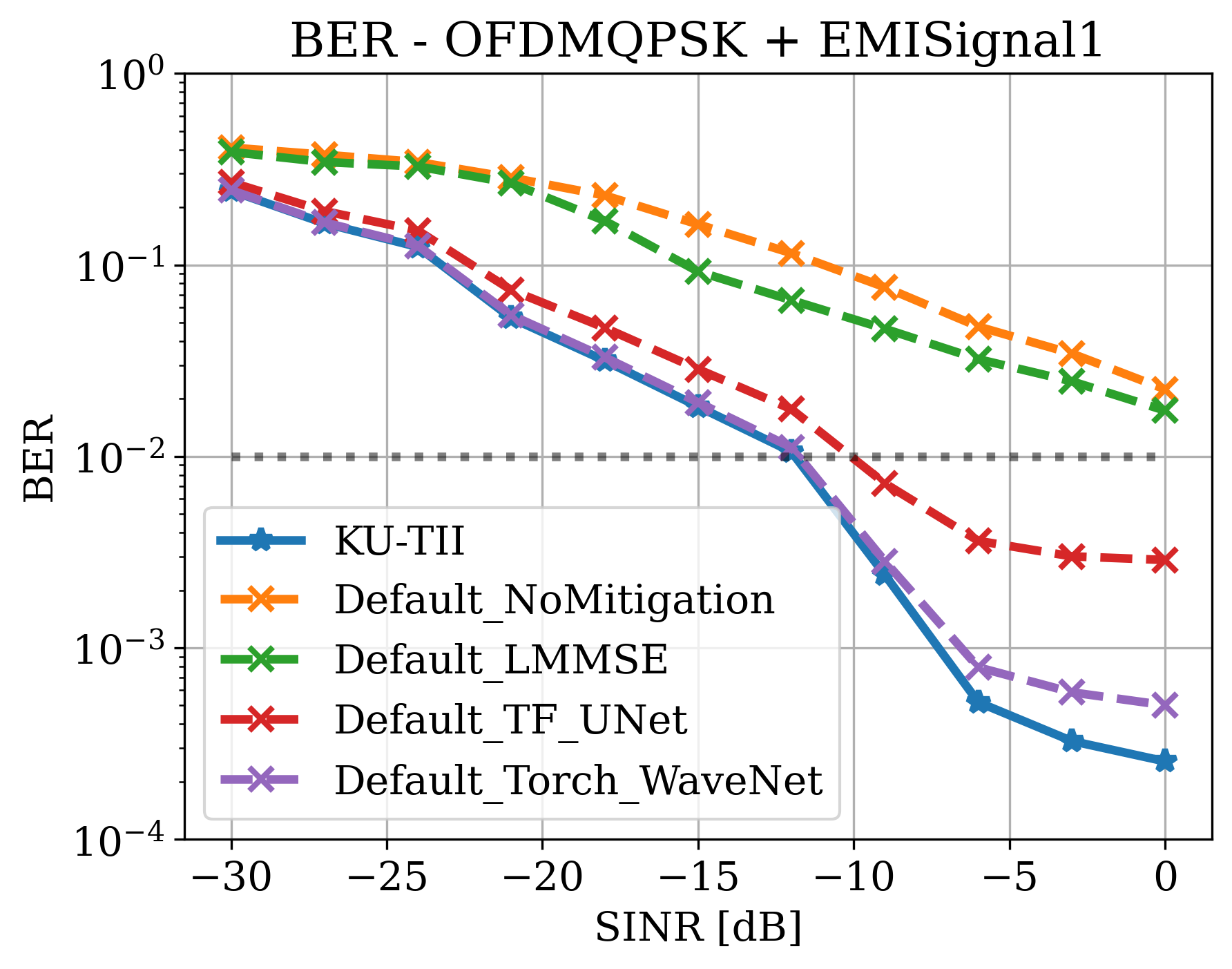}
        \caption{BER performance for OFDM-QPSK with EMISignal1}
        \label{fig:sub1}
    \end{subfigure}
    \quad
    \begin{subfigure}[b]{0.45\linewidth}
        \includegraphics[width=\linewidth]{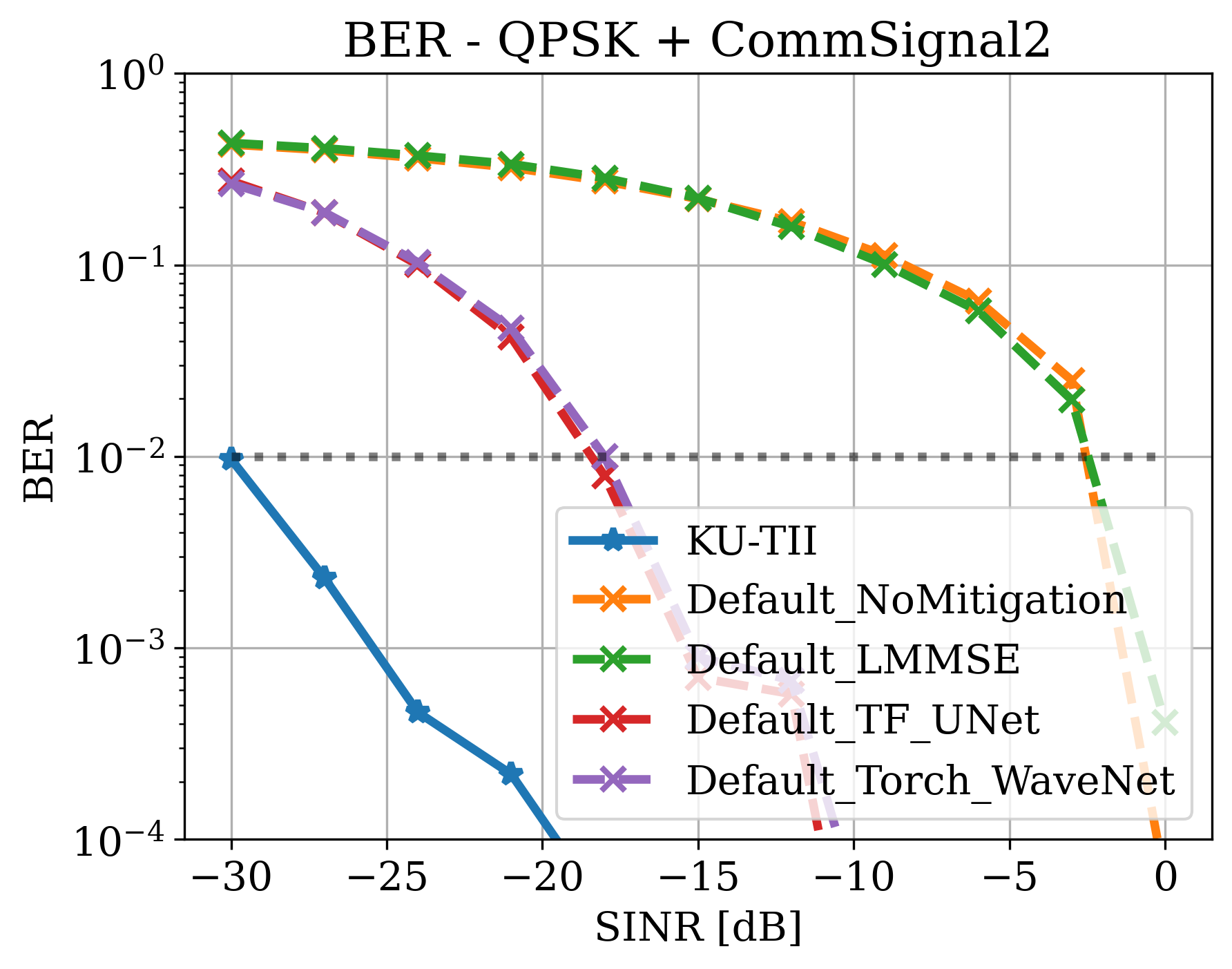}
        \caption{BER performance for QPSK with CommSignal2}
        \label{fig:sub2}
    \end{subfigure}
    \newline
    \begin{subfigure}[b]{0.45\linewidth}
        \includegraphics[width=\linewidth]{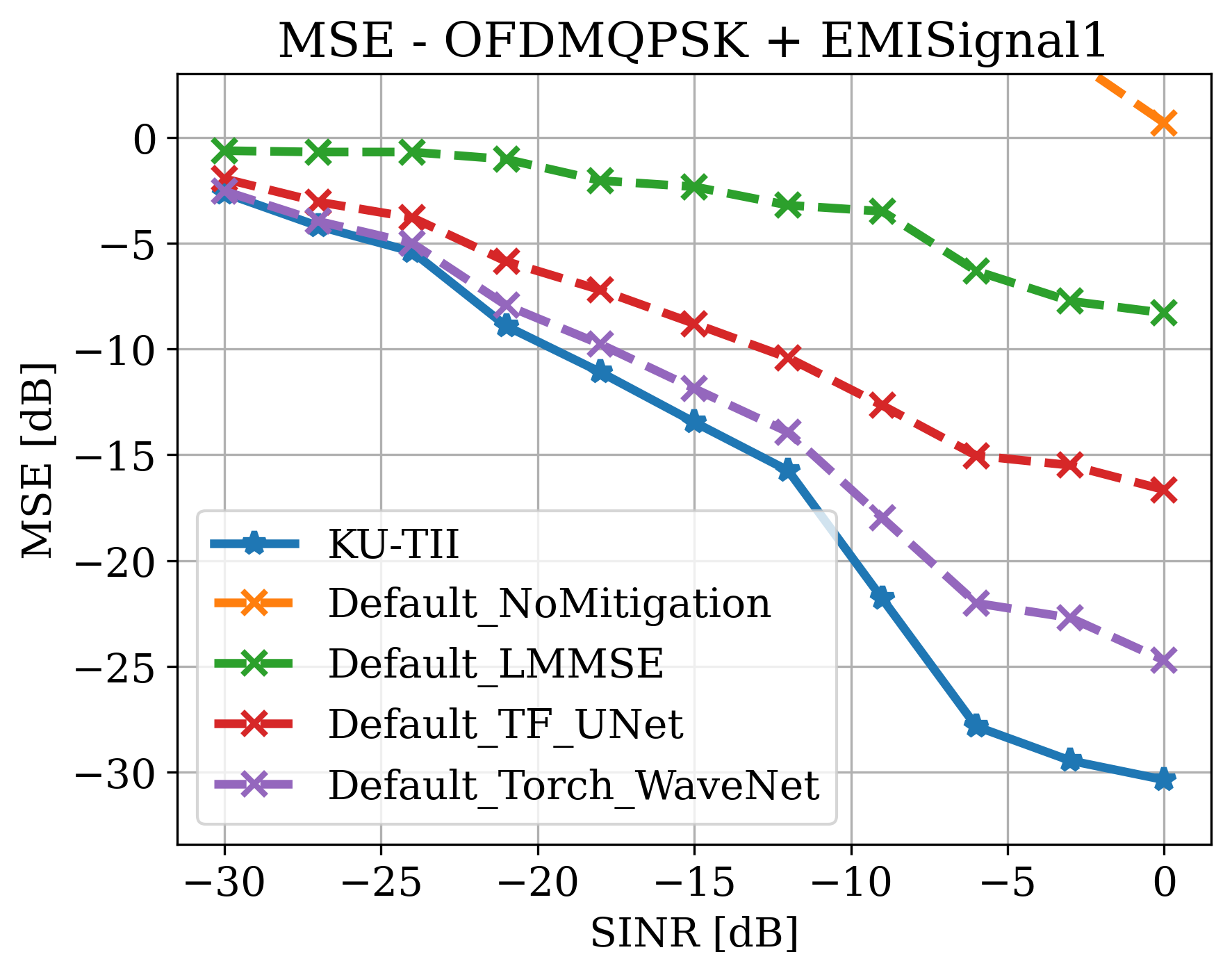}
        \caption{MSE performance for OFDM-QPSK with EMISignal1}
        \label{fig:sub3}
    \end{subfigure}
    \quad
    \begin{subfigure}[b]{0.45\linewidth}
        \includegraphics[width=\linewidth]{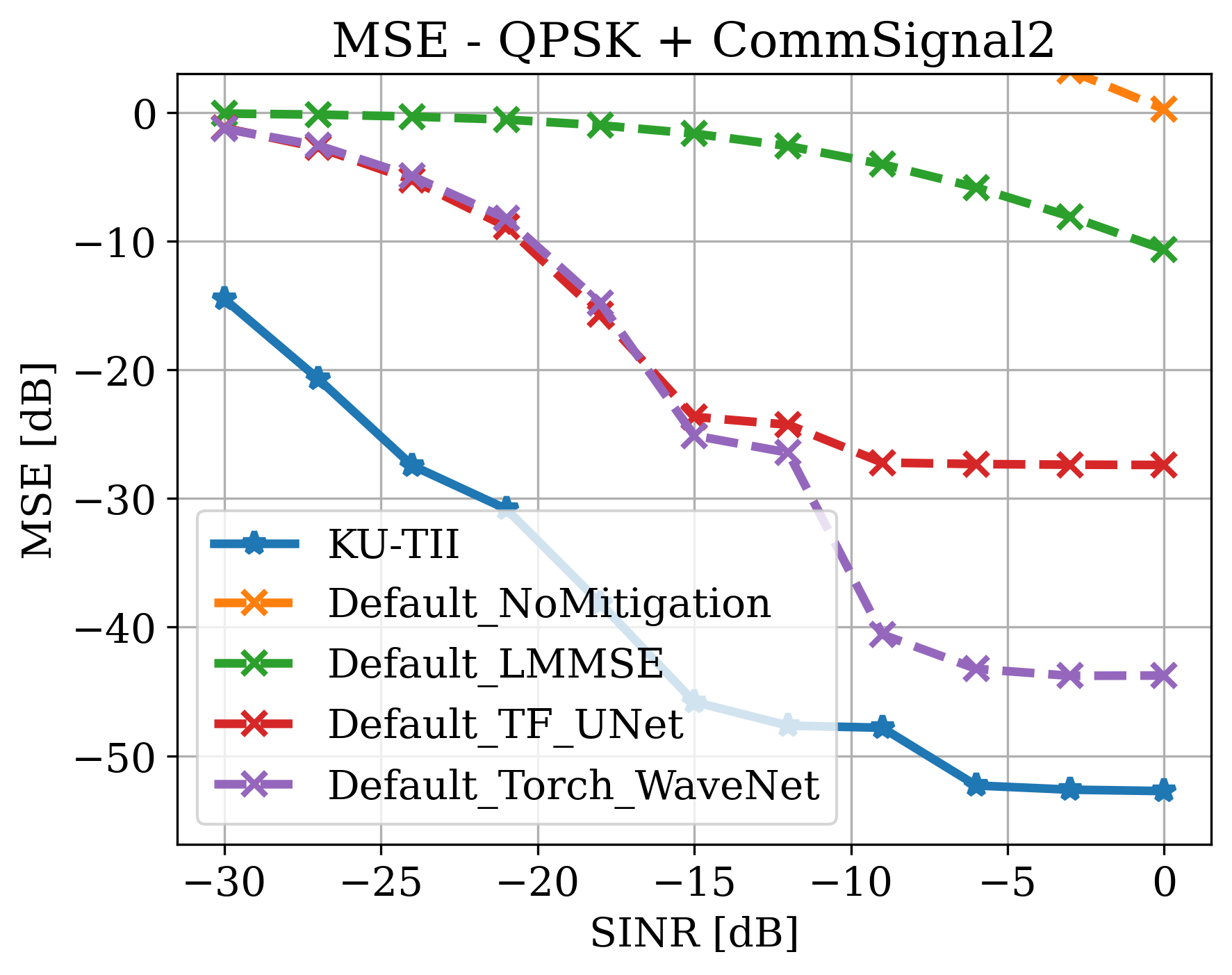}
        \caption{MSE performance for QPSK with CommSignal2}
        \label{fig:sub4}
    \end{subfigure}
    \caption{Comparative Performance Analysis of our Proposed Method Vs. Different Methods}
    \label{fig:composite}
    \vspace{-2mm}
\end{figure}

\section{Conclusion}
\label{sec:typestyle}
In this paper, we propose a novel WaveNet architecture with learnable dilation, which has yielded marked improvements in RF signal separation. Our tailored modifications and data augmentation methods have led to enhanced model performance. The results have demonstrated the effectiveness of our approach, potentially advancing the field of RF communications through Deep learning. Moving forward, we aim to investigate the scalability of our model to more complex signal environments and the integration of real-time adaptation capabilities, further solidifying its practical applications in the evolving landscape of RF communications.

\bibliographystyle{IEEEbib}
\bibliography{strings,refs}

\end{document}